\documentclass[aps,prl,showpacs,twocolumn,superscriptaddress,floatfix]{revtex4-1}

\usepackage[colorlinks,bookmarks=false,citecolor=blue,linkcolor=red,urlcolor=blue]{hyperref}
\usepackage{epsfig}
\usepackage{tikz}
\usepackage{graphicx}

\usepackage{dcolumn}
\usepackage{bm}

\usepackage{amsmath,amssymb}

\begin{document}

\title{Carrier driven antiferromagnetism and exchange-bias 
in SrRuO$_3$/CaRuO$_3$ heterostructures}
\author{Parul Pandey}\email{p.pandey@hzdr.de}
\affiliation{Helmholtz-Zentrum-Dresden-Rossendorf, Institute of Ion Beam Physics and Materials Research, Bautzner Landstra{\ss}e 400, 01328 Dresden, Germany}
\author{Ching-Hao Chang}\email{c.h.chang@ifw-dresden.de}
\affiliation{Leibniz-Institute for Solid State and Materials Research, Helmholtzstra{\ss}e 20, 01069� Dresden, Germany}
\author{Angus Huang}
\affiliation{Department of Physics, National Tsing Hua University, Hsinchu 30043, Taiwan}
\author{Rakesh Rana}
\affiliation{Helmholtz-Zentrum-Dresden-Rossendorf, Institute of Ion Beam Physics and Materials Research, Bautzner Landstra{\ss}e 400, 01328 Dresden, Germany}
\author{Changan Wang}
\affiliation{Helmholtz-Zentrum-Dresden-Rossendorf, Institute of Ion Beam Physics and Materials Research, Bautzner Landstra{\ss}e 400, 01328 Dresden, Germany}
\affiliation{Technische Universit\"at Dresden, 01062 Dresden, Germany}
\author{Chi Xu}
\affiliation{Helmholtz-Zentrum-Dresden-Rossendorf, Institute of Ion Beam Physics and Materials Research, Bautzner Landstra{\ss}e 400, 01328 Dresden, Germany}
\affiliation{Technische Universit\"at Dresden, 01062 Dresden, Germany}
\author{Horng-Tay Jeng}
\affiliation{Department of Physics, National Tsing Hua University, Hsinchu 30043, Taiwan}
\affiliation{Institute of Physics, Academia Sinica, Taipei 11529, Taiwan}
\author{Manfred Helm}
\affiliation{Helmholtz-Zentrum-Dresden-Rossendorf, Institute of Ion Beam Physics and Materials Research, Bautzner Landstra{\ss}e 400, 01328 Dresden, Germany}
\affiliation{Technische Universit\"at Dresden, 01062 Dresden, Germany}
\author{R. Ganesh}
\affiliation{The Institute of Mathematical Sciences, HBNI, C I T Campus, Chennai 600113, India}
\author{Shengqiang Zhou}
\affiliation{Helmholtz-Zentrum-Dresden-Rossendorf, Institute of Ion Beam Physics and Materials Research, Bautzner Landstra{\ss}e 400, 01328 Dresden, Germany}

\date{\today}

\begin{abstract}
Oxide heterostructures exhibit a rich variety of magnetic and transport properties which arise due to contact at an interface. This can lead to surprising effects that are very different from the bulk properties of the materials involved.  
We report the magnetic properties of bilayers of SrRuO$_3$, a well known ferromagnet, and CaRuO$_3$, which is nominally a paramagnet. We find intriguing features that are consistent with CaRuO$_3$ developing dual magnetic character, with both a net moment as well as antiferromagnetic order. 
We argue the ordered SrRuO$_3$ layer induces an undulating polarization profile in the conduction electrons of CaRuO$_3$, by a mechanism akin to Friedel oscillations. 
At low temperatures, this oscillating polarization is inherited by rigid local moments within CaRuO$_3$, leading to a robust exchange bias. 
We present \textit{ab initio} simulations in support of this picture. Our results demonstrate a new ordering mechanism and throw light on the magnetic character of CaRuO$_3$.

\end{abstract}
%\pacs{}
\maketitle 

\paragraph{Introduction:}
Atomically sharp heterostructures of correlated oxides host a rich variety of electronic and magnetic phases\cite{Hwang2012,Ngai2014,Mannhart2014}. 
Their study has lead to the discovery of several new principles and mechanisms such as Oxygen vacancy-driven order\cite{Jang2017}, exchange bias\cite{Nogues1999}, interfacial conduction\cite{Ohtomo2004,Thiel2006}, carrier-driven coupling\cite{Chang2017}, etc. 
In particular, heterostructures can serve as a sensitive probe by using proximity to an ordered system to bring out latent ordering in an otherwise disordered material. 
Here, we study a bilayer system composed of SrRuO$_3$ (SRO), a ferromagnetic metal, and CaRuO$_3$ (CRO), a paramagnetic metal. Contact with a ferromagnet brings out a complex magnetism in CRO, providing a window into its microscopic character. The bilayer system exhibits a robust exchange bias, a phenomenon with promise for technological applications.

SRO and CRO are well studied members of Ruthenate family. They have strong structural similarities but radically different magnetic properties. While SRO is a ferromagnetic bad metal\cite{Koster2012}, CRO is known to be a paramagnet on the verge of a magnetic transition\cite{Gunasekera2015}. This striking difference is the subject of a long running debate\cite{Longo1968,Yoshimura1999,Felner2003,Cao2008,Cheng2013,Dang2015}. It is presumably due to a small difference in the Ru-O-Ru bond angle, which itself arises due to the smaller ionic size of Calcium vis-\`a-vis Strontium. Although CRO is paramagnetic, studies have indicated that strain can induce ferromagnetism or antiferromagnetism in CRO\cite{Longo1968,Yoshimura1999,He2001,Tripathi2014}. Given the debate on the differing behaviour in CRO and SRO and the many conflicting results in literature, it is interesting to study bilayers with these materials in immediate contact. 
A nuanced picture emerges with magnetism in CRO driven by the SRO moment and mediated by conduction electrons. 

%\paragraph{Exchange bias:}
A key finding in this work is the occurence of exchange bias in a bilayer composed of a ferromagnet and a nominal paramagnet. Exchange bias\cite{Meiklejohn1956,Nogues1999} is a well known phenomenon that is best exemplified in a ferromagnet-antiferromagnet bilayer\cite{Panagiotopoulos1999}. When an applied field is varied, the antiferromagnet does not respond as it has zero net moment. As a result, the ferromagnetic layer remains in contact with a polarized system (the nearest atomic layer of the antiferromagnet). The exchange coupling to this layer adds to the effective field experienced by the ferromagnet, resulting in a shift of the M-H hysteresis curve opposite to the cooling field direction. This phenomenon also occurs when both layers are ferromagnetic\cite{Ke2005}; in this case, it arises from strong anisotropy in one layer which makes it insensitive to an applied field, thereby imposing an effective exchange field on the other layer. 
Heterostructures with an antiferromagnetic insulator (CaMnO$_3$ or Eu$_{0.42}$Sr$_{0.58}$MnO$_3$) and a paramagnetic metal (CaRuO$_3$)\cite{Nanda2007, He2012, Parul2015} also show exchange bias. In these cases, leakage of electrons from the metallic side effectively dopes a narrow window of the Manganite layer. This is believed to drive ordering, with the resulting ferromagnet-antiferromagnet interface leading to exchange bias. 
In this letter, we study SRO-CRO bilayers in which both layers are metallic. As there is no underlying doping-driven ordering (\`a la double-exchange) at play here, the underlying physics is very different.

\paragraph{Summary of key results:}
We present a detailed experimental study of SRO-CRO bilayers with various relative widths. 
Our central results are: (a) the bilayer system develops a net ferromagnetic moment, (b) in all samples, we find a single critical temperature which is close to the bulk T$_c$ of SRO,
(c) immediately below T$_c$, the strength of the net moment is directly correlated with CRO thickness; in contrast, at low temperatures, it is inversely correlated, (d) a large exchange bias is seen at low temperatures indicating insensitivity to an applied field, and (e) the behaviour of the magnetic moment is reflected in transport measurements indicating that conduction electrons play an important role. Based on these results, we argue that the CRO layer develops an oscillating polarization with two contributions: local moments and conduction electrons.
The conduction electrons develop Friedel-like oscillations in polarization due to the boundary condition imposed by the ordered SRO layer. This is inherited by rigid local moments at low temperatures. We present a simple theoretical model and ab initio simulations to support our picture.

\paragraph{Experimental results:}
High-quality SRO$_m$CRO$_n$ bilayers of various thickness were synthesized on an SrTiO$_3$ (100) substrate, where subscripts $m$ and $n$ denote the thicknesses of layers in nanometres. The SRO and CRO layers were deposited by the pulsed laser deposition technique, using a 248 nm Excimer laser at 680$^\circ$ C in 15 Pa of O$_2$ and 700$^\circ$ C in 30 Pa of O$_2$, respectively. Magnetization measurements were performed on a Quantum Design SQUID magnetometer with field applied in the in-plane direction. Resistivity was measured in a LakeShore HMS 9709A system.

We first discuss temperature dependence of the magnetization for samples with different relative widths, shown in Fig.~1(a). For all samples, magnetization was measured after cooling in a field of 100 Oe. We see that SRO$_{10}$CRO$_n$ (n = 4, 6, 8 nm), and SRO$_7$CRO$_4$ bilayers exhibit a single Curie-like temperature at 
$\sim$150 K, close to the bulk T$_c$ of SRO, 160 K\cite{Cao1997}. The marginal decrease in T$_{c}$, vis-\`a-vis bulk SRO, can possibly be due to strain in the thin film\cite{Rana2014}. In our thinnest sample, the SRO$_5$CRO$_4$ bilayer, we find a lower T$_c$ of $\sim$130 K. This sample also differs in being an insulator at low temperatures, possibly due to confinement effects\cite{Toyota2005,Toyota2006,Xu2016}.
As we decrease the temperature, the magnetization grows in all samples. We see a saturation, or even a small decrease, below $\sim$50K. On the whole, we emphasize that all bilayers show a single Curie-like temperature below which a moment develops. The proximity of this  temperature to the bulk T$_c$ of SRO indicates that it is the ordered moment of SRO which drives the magnetization of the bilayer.

Figs.~1(c,d) shows magnetic-isotherms of different samples, measured at 5 K. 
The saturation magnetic moment and the coercivity systematically increase upon decreasing the relative thickness of the CRO layer. This can be seen in the samples with an SRO layer of 5 to 10 nm width, where the CRO thickness is fixed at 4 nm. Likewise, when the SRO thickness is kept fixed at ~10 nm and CRO thickness is increased from 4 nm to 8 nm, the saturation moment and the coercivity decrease. This demonstrates that the net moment of the bilayer is \textit{inversely} correlated with CRO width at 5K. A very different picture emerges at high temperatures, just below T$_c$. As shown in Fig.~1(b), the net moment at 140 K increases when CRO relative width increases. Thus, at high temperatures, the net moment is \textit{directly} correlated with CRO width.

The features seen here are also reflected in transport measurements, with a peak in the magnetoresistance at T$_c$ (see Supplementary Materials).
The observed M-T and M-H curves show that the net bilayer moment depends on CRO thickness, indicating that CRO orders ferromagnetically. However, the absence of a second sharp feature (e.g., a second Curie-like temperature) in all samples rules out a magnetic transition in CRO. We note that the M-H curve shows a strong anisotropy with a sharp jump seen in the out-of-plane measurements (see Supplementary Materials). This is not surprising as both CRO and SRO are known to have strong uniaxial anisotropy\cite{Schultz2006,Koster2012}. 
 
\begin{figure}
\includegraphics[width=3.5in]{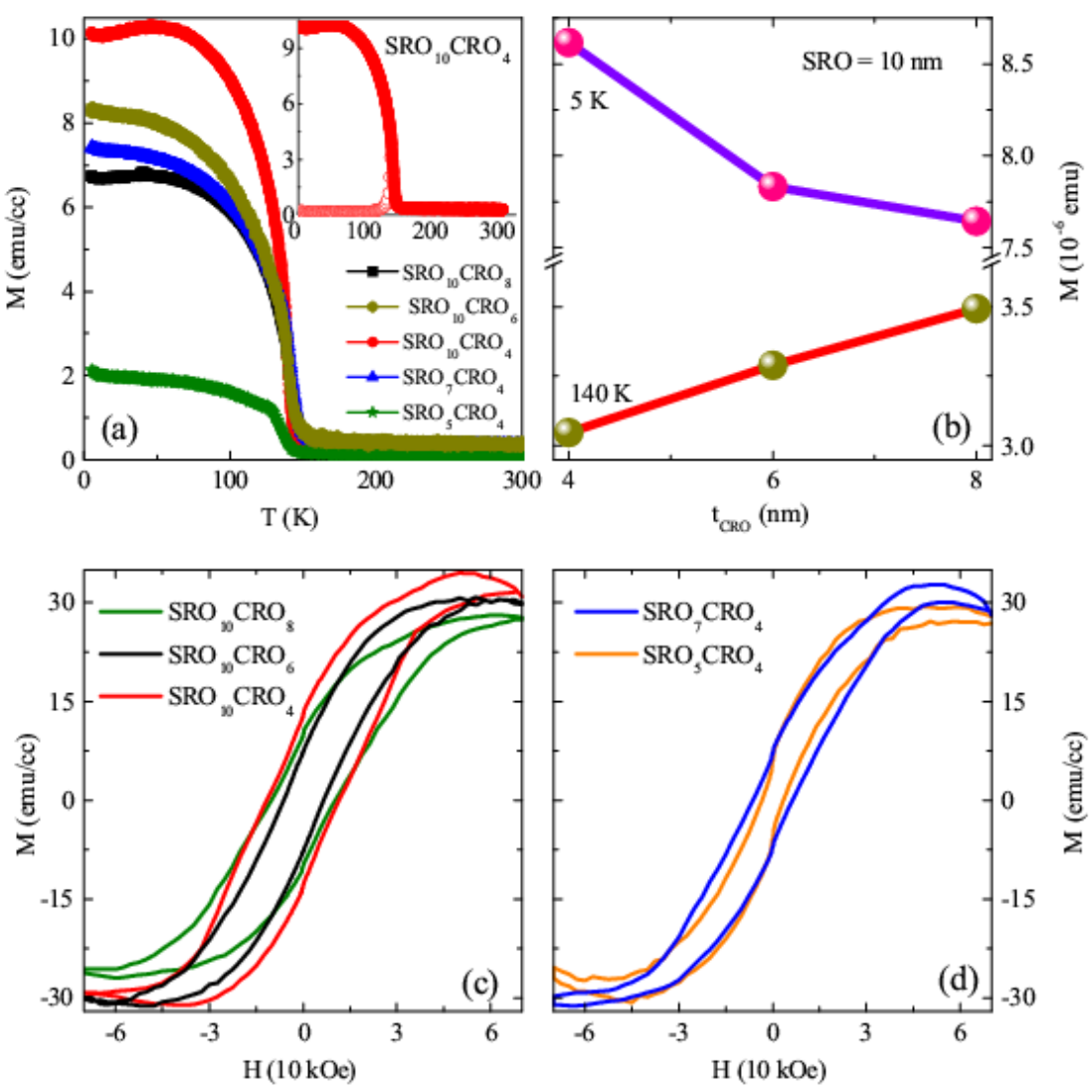}
\caption{Magnetization properties of the SRO-CRO bilayer. (a) Magnetization vs. temperature for samples with different layer widths. (b) Saturation moment vs. CRO width, with SRO layer width fixed at 10 nm, at low temperature (5K) and high temperature (140 K).
 (c \& d) Magnetization vs. applied field, measured at 5 K, for different layer widths. }
\label{fig.MT}
\end{figure}

\begin{figure*}
\includegraphics[width=6.5in]{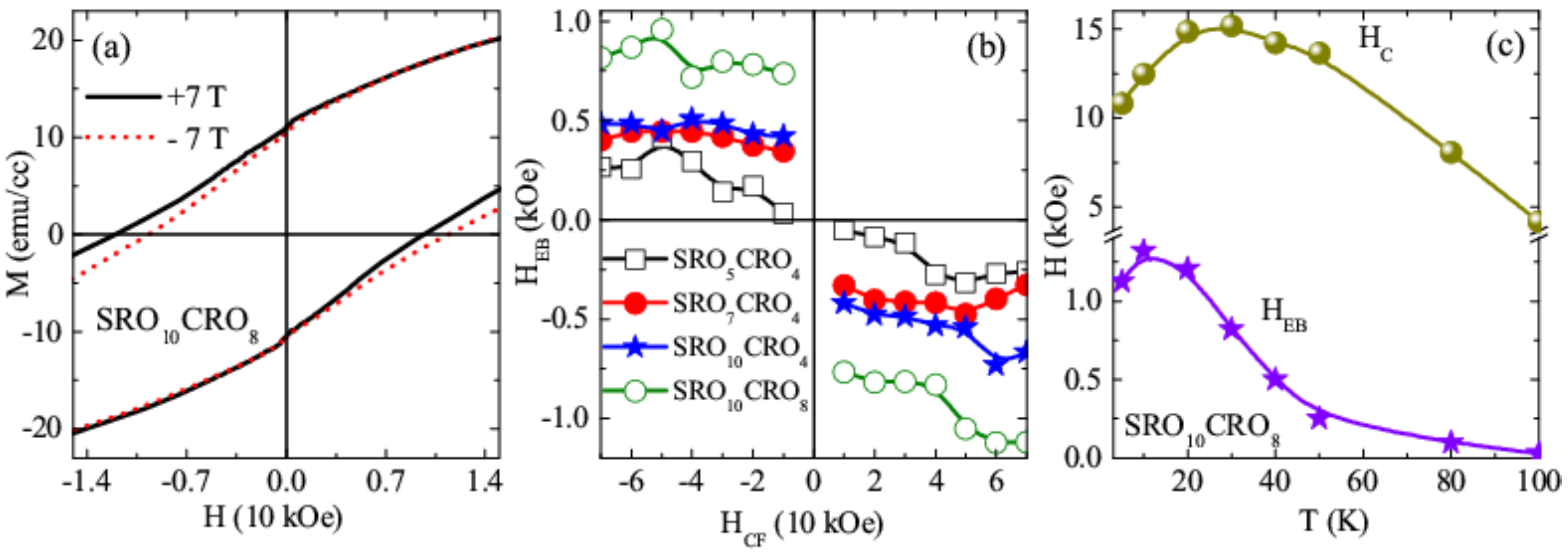}
\caption{Characterizing the M vs. H behaviour: (a) M-H curves for two different cooling fields, $\pm$ 7T, showing a shift of the hysteresis loop in the field direction. This signifies a significant exchange bias dependent upon the direction of the cooling field. (b) The obtained exchange bias (measured at 5K) as a function of cooling field for different layer widths. (c) Dependence of the exchange bias ($H_{EB}$) and the coercive field ($H_C$) upon temperature for a representative sample. }
\label{fig.MH}
\end{figure*} 
The M-H curves show clear exchange bias (EB). As shown in Fig. 2(a), the hysteresis loop shifts in the direction of the cooling field (see Supplementary Materials for more details). This is shown for  SRO$_{10}$CRO$_8$ bilayer at cooling fields of $\pm$7 T. 
This indicates a robust magnetic moment in the CRO layer which does not change with the applied field. Furthermore, this acts as a biasing field which favours ordering of the SRO moment along the cooling field. In Fig.~2(b,c), we plot the exchange bias field $H_{EB} = (H_1-H_2)/2$ and the coercive field $H_C = (H_1+H_2)/2$, where $H_1$ and $H_2$ are the positive and negative fields at which the magnetization vanishes. The coercive field $H_C$ changes smoothly with temperature, showing a small dip below $\sim$25 K.
The exchange bias shows strong temperature dependence; as shown in Fig.~2(c), it increases upon lowering the temperature. 
It is roughly independent of the strength of the cooling field as shown in Fig.~2(b).

\paragraph{Proposed mechanism:}
We first recapitulate essential features of SRO and CRO before presenting a scenario to explain the experimental findings. Both SRO and CRO are metallic at all temperatures (except for very thin layer widths); for the SRO-CRO bilayer, this dictates that conduction electrons can freely move throughout the system. As a consequence, interface-based coupling mechanisms such as superexchange are not applicable here\cite{Chang2017}. With SRO being a `bad metal', several studies suggest the existence of local moments which couple strongly with conduction electrons\cite{Jakobi2011,Kim2015,Dang2015}.
This is tied to a special property of SRO: its `minority carrier' nature -- conduction electrons are polarized \textit{opposite} to the net moment of SRO. This can be understood as follows. The Ru atoms in SRO are in the 3d$^4$ configuration with all four electrons occupying t$_{2g}$ orbitals At the simplest level, each Ru atom can be taken to be in the high spin configuration with 3 `up' electrons and one `down' electron. In the ferromagnetically ordered state, the `up' electrons are immobile due to Pauli blocking whereas the `down' spins can delocalize, leading to minority carrier conduction. The up-electrons can now be thought to form local moments while down-electrons form conducting bands. As CRO has essentially the same electronic structure as SRO, it has same the orbital configuration. This suggests that if magnetic order develops in CRO, it will have contributions from local moments as well as conduction electrons. Morever, the local moments and conduction electrons will have opposite moments.

To explain the magnetism seen in SRO-CRO bilayers, we argue that the magnetism in CRO occurs at two levels: conduction electrons and local moments. These two components respond to external magnetic fields at different time scales. While conduction electrons respond very quickly, the local moments are rigid. 
At high temperatures, above $T_c^{SRO}$, both SRO and CRO layers are not ordered. 
Immediately below $T_c^{SRO}$, the SRO layer develops an ordered moment. This strongly affects the conduction electrons of CRO in the form of a spin-dependent barrier potential. While one spin species can delocalize into the SRO layer, the other cannot. 
We build a minimal quantum well model to describe this effect. 

\begin{figure}
\includegraphics[width=3.25in]{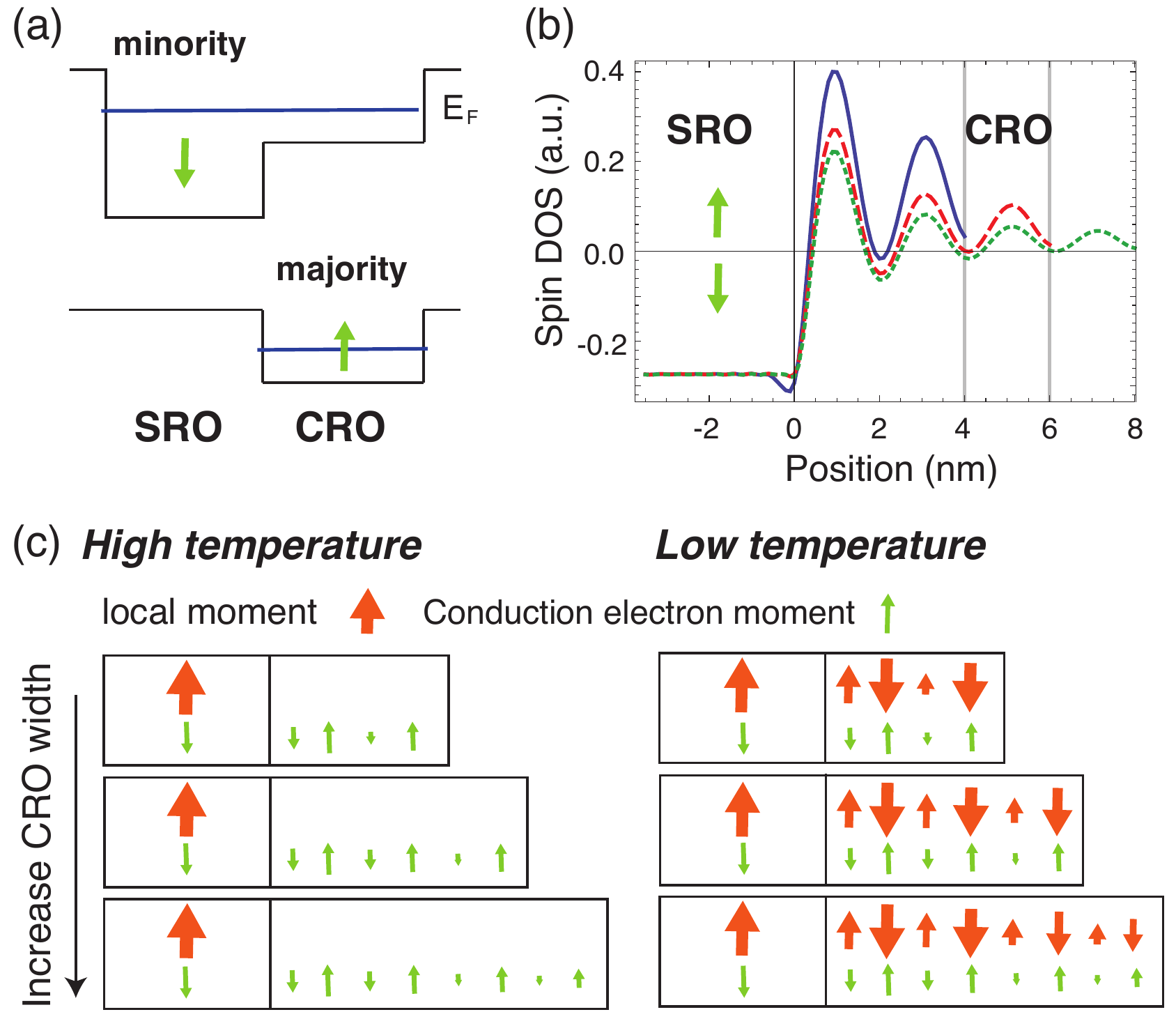}
\caption{Quantum well model. (a) Potentials seen by conduction electrons of the two spin species within the bilayer, assuming SRO to have ordered in the up-direction. (b) Obtained spin polarization (difference in DOS at the Fermi level between the two spin species) as a function of position, with $x>0$ corresponding to the CRO layer.
The three lines correspond to CRO widths of 4, 6 and 8 nm, with SRO width fixed at 10 nm. The CRO layer has an oscillating moment as well as a net moment in the up direction. (c) Nature of magnetic moments at high and low temperatures (see text). Local moments are shown as red arrows, while conduction electron polarization is shown in green. }
\label{fig.quantumwell}
\end{figure}

As shown in Fig.~\ref{fig.quantumwell}(a), we model the bilayer as an infinite slab (finite thickness in the $z$ direction, but infinite in the $x$ and $y$ directions). We take the conduction electrons to be free particles moving within this slab. Within the SRO layer, the two spin species see different $z$-dependent potentials. We take the potential to be $0.5$ eV ($-0.5$ eV) for the majority (minority) species, in accordance with band structure results for the Zeeman splitting\cite{abb15}. The  potential barriers which totally reflect electrons at the substrate and the surface are also set at $0.5$ eV. 
The effective mass of conduction electrons is taken to be 4.4 m$_{\rm e}$, where m$_{\rm e}$ is the electron rest mass, and the Fermi velocity to be $k_F= 1.5\times 10^{-9}~m^{-1}$. 
These values are close to those measured in a CRO film \cite{sch14}. 
We point out that such a low Fermi velocity as observed in CRO layers is special; it leads to an ultra-long Fermi wavelength $\lambda_F=2\pi/k_F\sim 2$ nm, with observable consequence in our systems with CRO layers of several nm.

Within this model, we calculate the spin-resolved density of states of conduction electrons at the Fermi level. We find a position-dependent polarization which oscillates as we move into the CRO layer. This is precisely a realization of Friedel oscillations, a well known phenomenon in metals wherein an impurity generates short ranged density oscillations with wavevector $2k_F$. This is a characteristic response of low energy electron-hole excitations near the Fermi surface. Here, the SRO layer is effectively a one-dimensional spin-dependent impurity. It generates Friedel oscillations within the CRO layer with different amplitudes in the two spin species. Effectively, this leads to an oscillating polarization profile as shown in Fig.~\ref{fig.quantumwell}(b). 

The induced polarization profile in CRO has dual character. On the one hand, it represents short ranged antiferromagnetic order due to the oscillating polarization. However, it also builds a net moment within the CRO layer -- as seen in Fig.~\ref{fig.quantumwell}(b), the average polarization is non-zero. This net moment is generically \textit{along} the SRO moment. At `high' temperatures (i.e., immediately below T$_c$), we argue that the conduction electrons in CRO acquire this polarization while the local moments have not yet ordered as shown in Fig.~\ref{fig.quantumwell}(c). As a consequence, the CRO layer \textit{adds} to the net moment of the bilayer.

\begin{figure}
\includegraphics[width=3.5in]{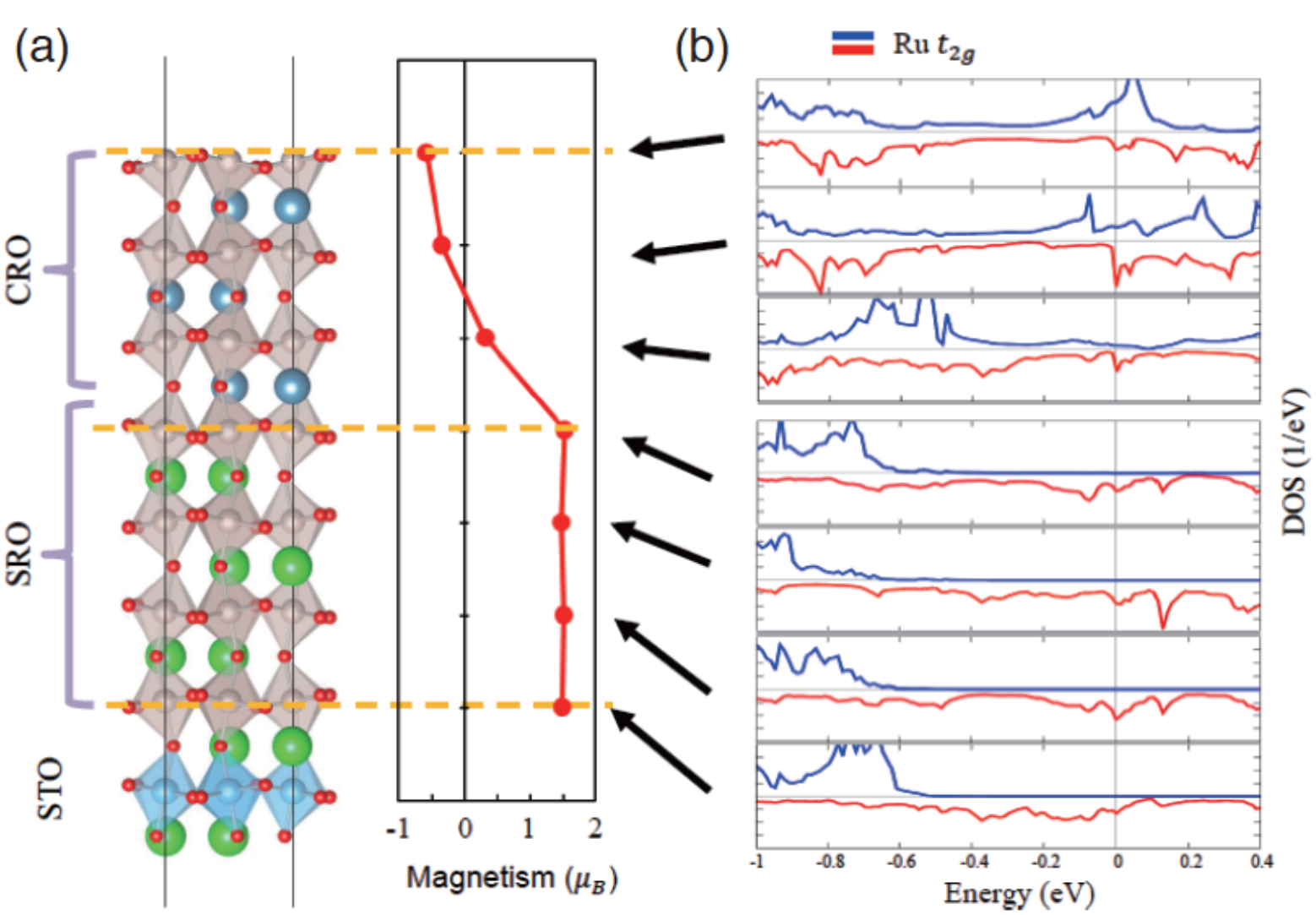}
\caption{\textit{Ab initio} simulations. (a) Crystal structure used for the SRO-CRO bilayer. The obtained magnetization from the simulations as a function of position within the heterostructure. (b) The spin-resolved density of states as a function of energy at different positions within the bilayer. The red line corresponds to spin which forms the minority species within SRO while the blue line is the majority species.  
 }
\label{fig.abinitio}
\end{figure} 
 
\paragraph{Role of field cooling in establishing exchange bias:}
Friedel-like oscillations in the conduction electrons of CRO cannot explain the observed exchange bias, which requires a rigid moment that is insensitive to an applied field. As conduction electrons are highly mobile, we expect that they will rearrange themselves very quickly to an applied field. Instead, rigid moments could arise from the local moment spins of CRO.  We argue that during the gradual field cooling process, the polarization profile of the conduction electrons is imprinted on local moments. As CRO is likely to be a minority carrier system, the local moments at any given point orient in the direction opposite to the conduction electron moment as shown in Fig.~\ref{fig.quantumwell}(c). Thus, at low temperatures, the local moments develop an oscillating polarization profile, with a net moment that is \textit{opposite} to the SRO moment. This explains how the net moment of the bilayer is directly correlated to CRO width at high temperatures (when only conduction electrons in CRO contribute) and inversely correlated at low temperatures (when local moments outweigh the conduction electrons).

We argue that unlike the conduction electrons, the local moments respond very slowly to an applied field. 
In addition, the polarization profile of the local moments has an oscillating component. This antiferromagnet-like character makes them insensitive to an applied field (as in conventional exchange bias in a ferromagnet-antiferromagnet bilayer). This picture is supported by the exchange bias data. We find that $H_{EB}$ increases with decreasing temperature as shown in Fig.~\ref{fig.MH}(c), becoming negligible in the vicinity of T$_c$. This is consistent with local moment order becoming progressively weaker with increasing temperature.

Our scenario for the ordering at low temperatures is supported by \textit{ab initio} results which show an oscillating polarization within CRO with a net moment that is opposite to that of SRO (see Supplementary Materials for details). As shown in Fig.~\ref{fig.abinitio}, we find a spatially varying moment with oscillations within the CRO layer. Fig.~\ref{fig.abinitio}(b) shows the spin-resolved density of states (DOS) vs. energy at different positions within the bilayer. In both the SRO and CRO regions, we find that the DOS at the Fermi level is \textit{anti-correlated} with the local moment. We can interpret this as follows -- the conduction electrons at the Fermi level are polarized opposite to the local magnetic moment at each point. Both conduction electrons and the local moment have an oscillating magnetization profile, consistent with Fig.~\ref{fig.quantumwell}(c).    

\paragraph{Discussion:}
We have presented evidence for a dual magnetism that emerges in CRO when placed in contact with SRO in a bilayer geometry. Our work suggests a new mechanism of ordering that is mediated by an oscillatory response of charge carriers. This principle could underlie the recent observation of exchange bias in bilayers of LaMnO$_3$ (an insulating ferromagnet in thin film form) and LaNiO$_3$ (a metallic paramagnet)\cite{Gibert2012}. The charge carriers in the nickelate develop an oscillatory behaviour with an antiferromagnetic character, consistent with \textit{ab initio} results\cite{Gibert2012}. As in our study, the exchange bias increases at low temperatures where Ni local moments may give rise to rigidity.

Our results shed light on the magnetism of CRO. More generally, our results show how interfaces can induce complex magnetization profiles mediated by conduction electrons. A complex interplay emerges among magnetization, exchange bias and transport properties. This suggests new ways to realize a tunable exchange bias, with potential for technological applications.  

\paragraph{Acknowlegements} 
P.P. acknowledges Deutsche Forschungsgemeinschaft (ZH 225/6-1), Germany for the financial support and Ilona Skorupa for providing the SRO target.
C.-H.C. acknowledges financial support from Deutsche Forschungsgemeinschaft, Grant CH 2051/1-1.
C.-H.C., A.H., and H.-T.J. acknowledge support from the National Center for Theoretical Sciences. R.G. thanks IFW Dresden for hospitality.
C.-H.C. thanks Ulrike Nitzsche for technical assistance.
%\bibliographystyle{apsrev4-1} 
%\bibliography{SRO_CRO}
%

\end{document}